\documentclass{ws-ijmpe}
\usepackage{graphicx,color,bm}

\def\eg{{\it e.g.,}}
\def\ie{{\it i.e.,}}



\begin{document}

\markboth{A. Baran, Z. \L ojewski and K. Sieja} {Pairing and
  $\alpha$-decay} 

\title{Pairing and $\alpha$-decay} 

\author{\footnotesize A. BARAN} \address{\it Institute of Physics,
  University of M. Curie-Sk\l odowska,\\ ul. Radziszewskiego 10,
  20-031 Lublin, Poland}

\author{\footnotesize Z. \L OJEWSKI} \address{\it Institute of
  Informatics, University of M. Curie-Sk\l odowska,,\\ Pl.
  M.~Curie-Sk³odowskiej 1, 20-031 Lublin, Poland}

\author{\footnotesize K. SIEJA} \address{\it Institute of Physics,
  University of M. Curie-Sk\l odowska,\\ ul. Radziszewskiego 10,
  20-031 Lublin, Poland\\ \& \\Centre d'Etudes Nucl{\'e}aires de
  Bordeaux Gradignan, Le Haut Vigneau BP120, F-33175 Gradignan,
  France}

\date{~}
\maketitle

\begin{history}
\received{~}
\revised{~}
\end{history}


\maketitle

\begin{abstract}
  Nuclear pairing interaction plays a crucial role in both
  macroscopic-microscopic and fully macroscopic descriptions of
  nuclei. In the present study we discuss different pairing
  interactions (monopole and $\delta$ pairing forces) and the methods
  allowing for the particle number symmetry restoration in addition to
  the customary BCS treatment of pairing correlations in the context
  of $\alpha$-decay half-lives for superheavy nuclei. The calculations
  are done in the macroscopic-microscopic framework for even-even
  nuclei with $Z> 110$.

\end{abstract}

\section{Introduction}
In the macroscopic-microscopic treatment of nuclear energy one needs
to know the dependence of the total energy of the nucleus on
deformation parameters (\eg\ an elongation).  In this method the
energy is a sum of the macroscopic part ($E_{\rm macro}$) and
microscopic energy consisting of shell ($\delta E_{\rm shell}$) and
pairing ($\delta E_{\rm pair}$) energies
\begin{equation}
V({\rm def}) = E_{\rm macro} + \delta E_{\rm shell} + \delta E_{\rm pair}\,.
\label{eq-barrier}
\end{equation}

The influence of the type of macroscopic energy models on both fission
and on $\alpha$-decay was discussed recently.\cite{Baran05b,Baran05e}
For the macroscopic part we choose the Yukawa+Exponential model (YpE).
\cite{Krappe79} The shell energy depends only on the single particle
model -- here Woods-Saxon potential.\cite{Dudek79} The third component
of the total energy is the pairing energy customary obtained with the
monopole pairing force, \ie\ the pairing force with constant matrix
elements ($g=\rm const$) in the BCS model.  As it is known, the usual
BCS treatment of correlations leads to the particle number
non-conservation. To correct the solutions one may apply \eg\ exact
projection techniques, \ie\ variation after projection method or use
some approximative approaches like solving Lipkin-Nogami equations.
Both these methods are used here in the case of seniority pairing.

In the following a short description of Lipkin-Nogami (LN) model and a
variation after projection (VAP) method are presented together with
the consequences of both these techniques on $\alpha$-decay energies
($Q_\alpha$-values) and $\alpha$-half-lives ($T_\alpha$) in the region
of superheavy nuclei ($Z>110$).

\section{Pairing models\label{sec-models}}
In this chapter we give a short review of the pairing models which
will be used in our calculations. At each stage of the presentation we
will also show some intermediate results concerning pairing properties
of considered nuclei.

\subsection{State dependent $\delta$-pairing}
The $\delta$ type pairing interaction induces the state dependence of
the pairing gaps. In this case one has to solve a large number of BCS
equations in order to obtain all the gaps in the pairing window (in
this paper we solve $N+1$ equations for neutrons and $Z+1$ for protons).
The strength of the $\delta$-force which reproduces pairing gaps and
masses in superheavy region of nuclei is $225$ MeV fm$^3$ for both
neutrons and protons.\cite{Baran05e} In the case of monopole pairing
we have used $g$ parameters reported in Ref.\cite{Dudek80}
\begin{figure}
\begin{center}
\includegraphics[scale=0.475]{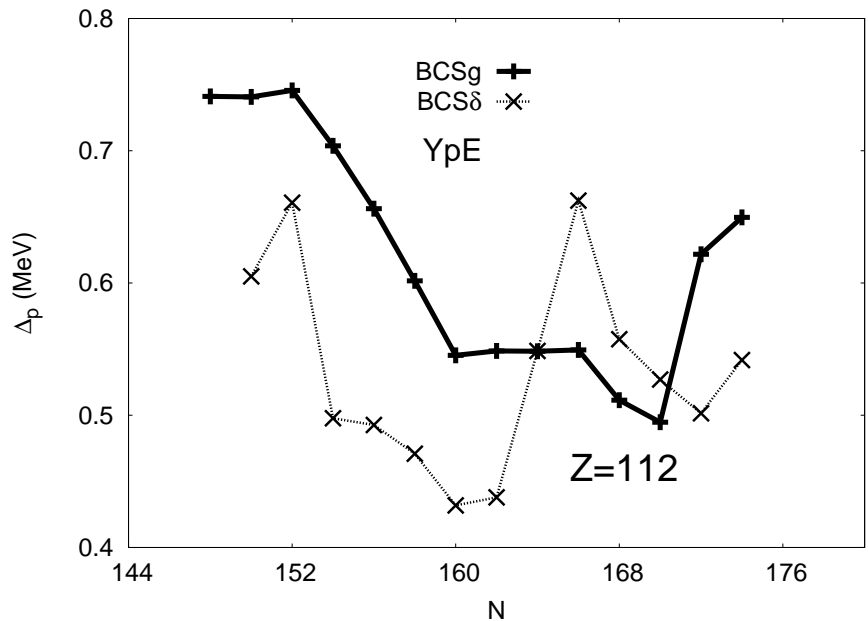}
\caption{Fermi pairing gaps in the case of BCSg and
  BCS$\delta$ for isotopes of Z=112.}
  \label{fig-dl-p-gd}
\end{center}
\end{figure}
\begin{figure}
\begin{center}
\includegraphics[scale=0.475]{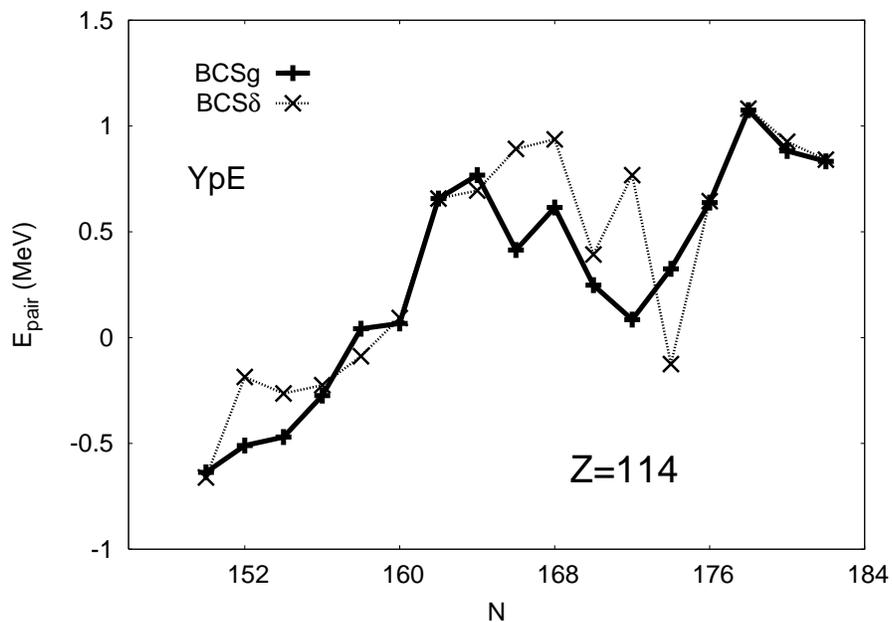}
\caption{Pairing corrections ($\delta E_{\rm shell}$) in the case of
  BCSg and BCS$\delta$ for isotopes of Z=114.}
  \label{fig-depair-gd}
\end{center}
\end{figure}

Figures \ref{fig-dl-p-gd} and \ref{fig-depair-gd} show the Fermi level
pairing gaps and pairing energies in both seniority BCS (BCSg) and
state dependent BCS (BCS$\delta$) models.  The differences in pairing
gaps are of the order of 0.1MeV. The deviations in the pairing energy
are in some cases larger than 0.5MeV.

\subsection{Lipkin-Nogami method }
In the case of Lipkin-Nogami (LN) model one reduces the fluctuations
of the particle number by adding to the Hamiltonian $\hat H$ the terms
$-\lambda\hat N$ and the quadratic term $-\lambda_2 (\hat
N-\langle\hat N\rangle)^2$ and one minimizes the average energy with
respect to $\lambda$. The procedure described in \eg\
Ref.\cite{Magierski93,Sieja07} gives the following expression for the
new corrected energy
\begin{equation}
E_{\rm LN} = E-\lambda_2\langle\Psi|\Delta\hat N|\Psi\rangle\,,
\label{eq:LNenergy}
\end{equation}
where
\begin{equation}
\lambda_2 = \frac{\langle\Psi |\hat H(\Delta\hat N)^2|\Psi \rangle}
                 {\langle\Psi |(\Delta\hat N)^2|\Psi\rangle}\,,
\qquad\Delta\hat N=\hat N-\langle\Psi|\hat N|\Psi\rangle\,, 
\end{equation}
$\hat N$ is the number operator and $\Psi$ is the BCS ground state.

\subsection{Variation After Projection approach}
Variation after projection (VAP) or Full BCS (FBCS) is an exact
particle number projection method \cite{Dietrich64,Sheikh00}.
Particle number projected wave function is
\begin{equation}
  \Psi=C\oint d\zeta\,\zeta^{-n_0-1}\prod_\nu(u_\nu+v_\nu\zeta 
  a^\dagger_\nu a^\dagger_{\bar\nu})\Phi_0\,,
\label{eq-proj}
\end{equation}
where $\Phi_0$ is the vacuum state for particles, $n_0$ the number of
nucleonic pairs, and $C$ the normalization constant
\begin{equation}
  \label{eq:6}
  |C|^2=1/(-4\pi^2)R_0^0\,,
\end{equation}
where $R_0^0$ is defined in Eq. (\ref{eq:2}).

The energy of the system of $N$ interacting fermions defined as
\begin{equation}
  \label{eq:8}
  E=\langle\Psi|\hat H|\Psi\rangle/\langle\Psi|\Psi\rangle\,,
\end{equation}
is given by
\begin{equation}
  \label{eq:1}
  E=\frac{1}{R^0_0}\left(\sum_{k>0}(2\epsilon_k-g_{kk})v_k^2 R_1^1(k)
    -\sum_{p,q; p\neq q} g_{pq} u_pv_p u_qv_q R^2_1(p,q) \right)\,.
\end{equation}
The $R_n^N(k_1,\dots,k_n)$ are the residues of contour integrals on
the complex plane
\begin{equation}
  \label{eq:2}
  R^N_n(k_1,\dots,k_N)=\frac{1}{2\pi i}\oint dz \frac{z^n}{z^{n_0+1}}
                      \prod_{k\neq k_1,\dots,k_N}(u_k^2+zv_k^2)\,.
\end{equation}
Here $v_k$ and $u_k$ are the parameters such that $u_k^2+v_k^2=1$ and
the number of pairs of nucleons reads
\begin{equation}
  \label{eq:7}
  n_0=\sum_k v_k^2R^1_1(k)/R^0_0\,.
\end{equation}
In order to calculate the residuum integrals we have used Ma and
Rasmussen recurrence relations.\cite{Ma77}

The FBCS equations are highly nonlinear and hard to solve even in the
case of seniority force $g_{kl}=\rm const$. They read
\begin{equation}
  \label{eq:3}
  2(e_k+\Lambda_k)u_kv_k-\Delta_k (u_k^2-v_k^2)=0\,,
\end{equation}
where
\begin{equation}
  \label{eq:4}
  e_k=(\epsilon_k-1/2g_{kk})\frac{R_1^1(k)}{R^0_0}\,,
\end{equation}
\begin{equation}
  \label{eq:4d}
  \Delta_k=\sum_lg_{kl}u_lv_l\frac{R^2_1(k,l)}{R^0_0}\,,
\end{equation}
and
\begin{eqnarray}
  \label{eq:5}
  \Lambda_k &=&-\,\frac{R_1^1(k)-R_0^1(k)}{2R_0^0} E 
  + \sum_{l>0}e_lv_l^2\frac{R_2^2(l,k)-R_1^2(l,k)}{R_0^0}\nonumber\\
  && \quad- \frac{1}{2}\sum_{lm}g_{lm}u_l v_l u_m v_m 
            \frac{R_2^3(k,l,m)-R_1^3(k,l,m)}{R_0^0}\,.
\end{eqnarray}

\section{$\alpha$ half-lives\label{sec-results}}
\begin{figure}
\begin{center}
\includegraphics[scale=0.475]{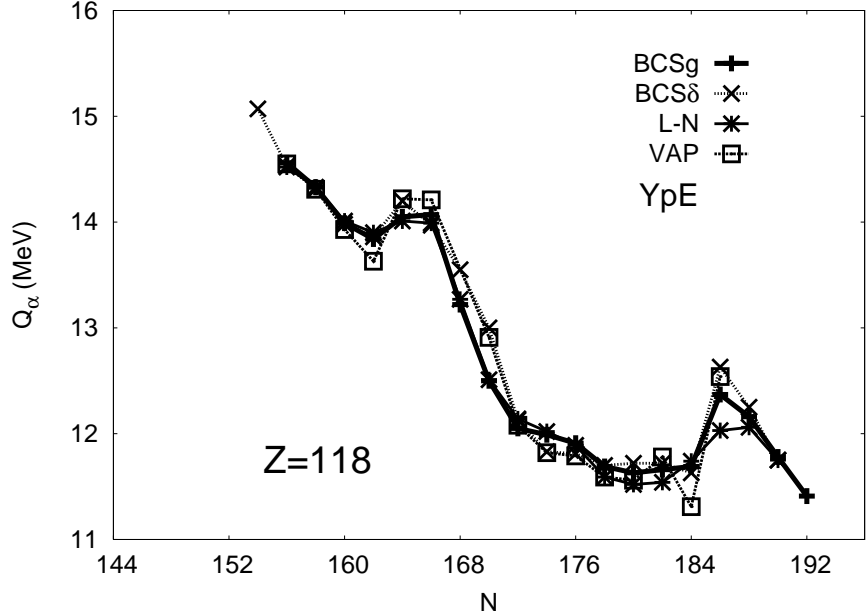}
\caption{$\alpha$-decay energies for all models for isotopes of $Z=118$.}
  \label{fig-qalpha-all}
\end{center}
\end{figure}
\begin{figure}
\begin{center}
\includegraphics[scale=0.475]{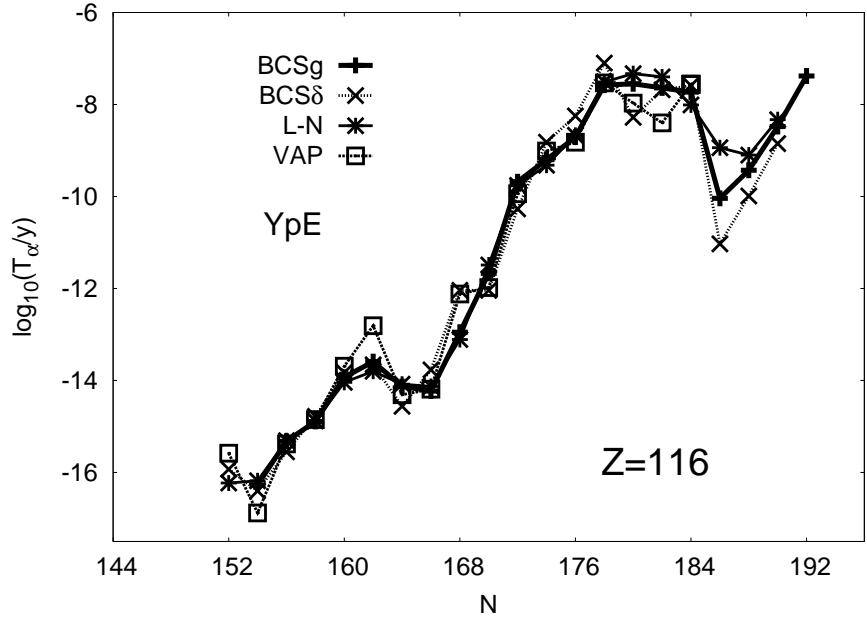}
\caption{Alpha half-lives for all models for isotopes of $Z=116$ isotopes.}
  \label{fig-logt-all}
\end{center}
\end{figure}
\begin{figure}
\begin{center}
\includegraphics[scale=0.475]{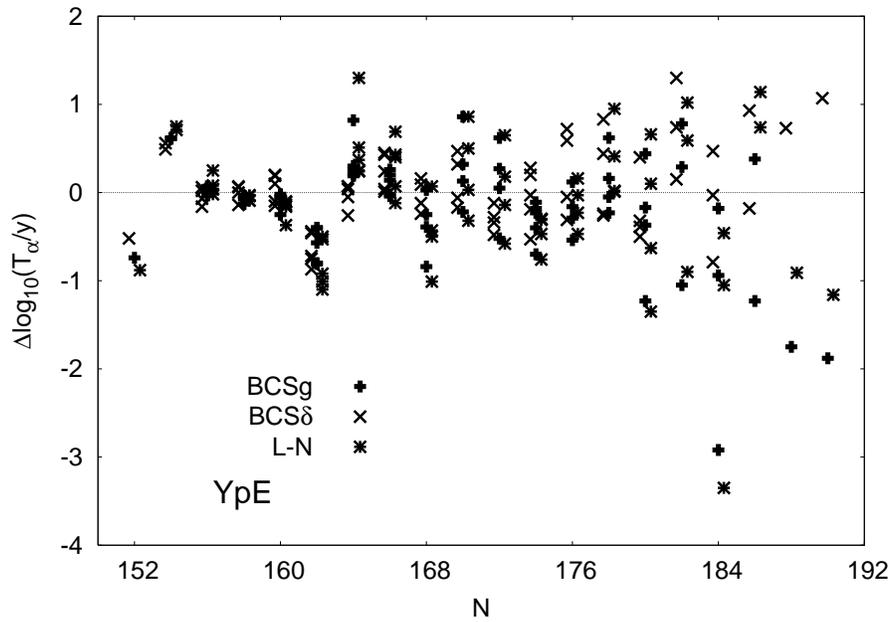}
\caption{The differences $\log(T_\alpha/y)-\log(T_\alpha/y)_{VAP}$
 versus neutron number $N$ for all models.}
  \label{fig-ta-dif}
\end{center}
\end{figure}
\begin{figure}
\begin{center}
\includegraphics[scale=0.475]{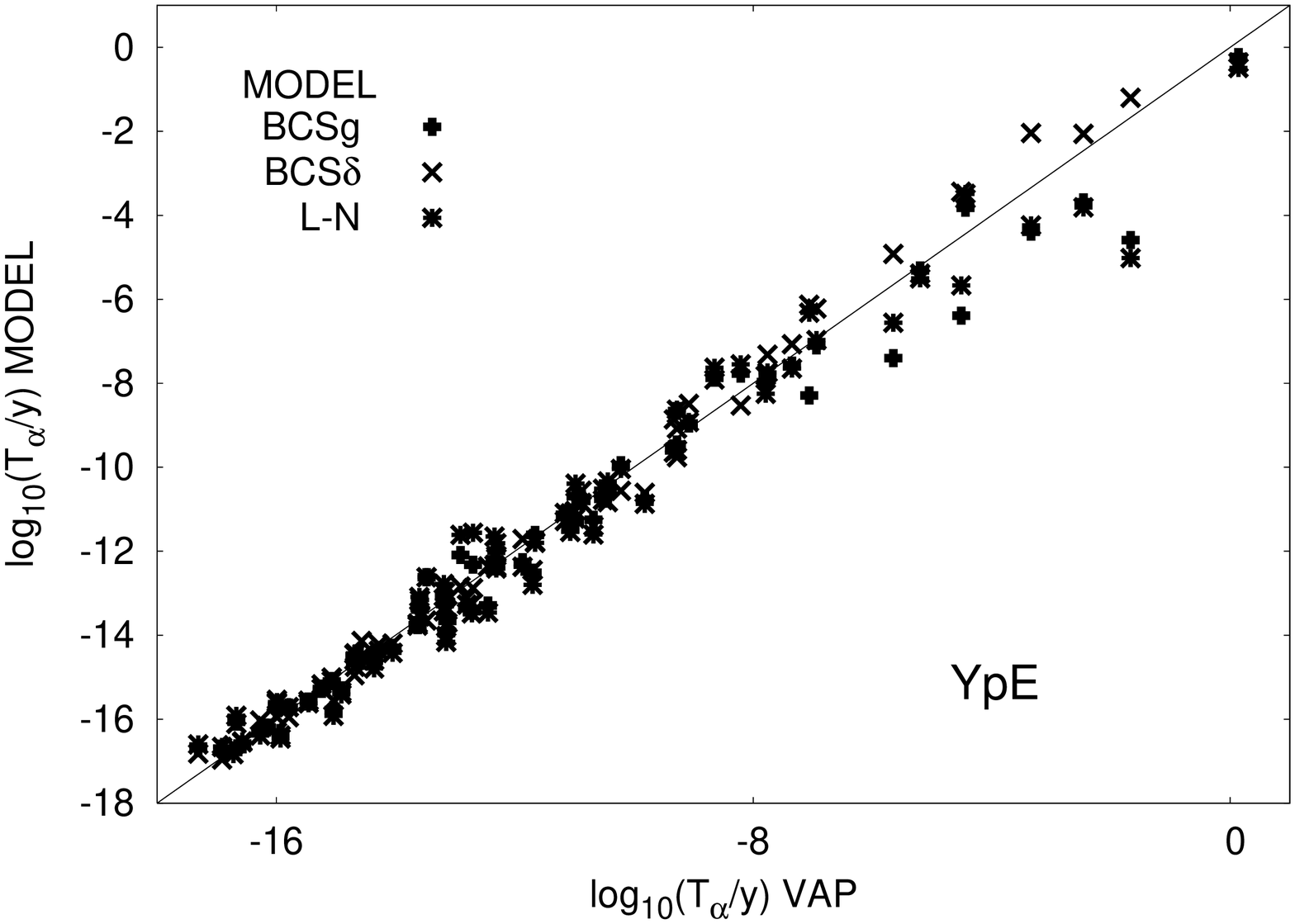}
\caption{Correlations of logarithms of half-lives $\log_{10}(T_\alpha)$
  relative to those from the VAP method (abscissa) for 70 nuclei
  with $112\le Z\le122$.}
  \label{fig-ta-cor}
\end{center}
\end{figure}
Both the approximate LN and exact VAP projection methods have been
applied to calculate macroscopic-microscopic masses, $\alpha$-decay
energies (Q-values) and half-lives ($T_\alpha$) for even-even isotopes
of superheavy nuclei: $112\le Z\le120$. The $\alpha$ half- lives are
determined from modified Viola-Seaborg formula.\cite{Patyk91} In the
following section we discuss the results of these calculations.

The $Q_\alpha$-values 
\begin{eqnarray}
 {Q}_{\alpha}(Z, N) =
  [ {\cal M}(Z, N)-{\cal M}(Z-2,N-2)-{\cal M}(2,2)  ] c^2 \,,
\end{eqnarray}
where ${\cal M}$ is the nuclear mass, are depicted in Figure
\ref{fig-qalpha-all} for all considered models. The pattern of
$Q_\alpha$ vs. neutron number $N$ is similar in all of the cases.
However, in the vicinity of $N=162$ and $N=184$ one observes some
discrepancies which will influence the calculations of
$\alpha$-half-lives.  This effect can be noticed on the Figure
\ref{fig-logt-all} where the logarithms of the half-lives (in years)
are displayed.  The largest differences occur in the vicinity of magic
numbers where the standard BCSg and BCS$\delta$ methods collapse
yielding no superfluid solutions.

Figure \ref{fig-ta-dif} shows the discrepancy between the half-lives
(logarithms) vs. the neutron number $N$ in the case of all of the
models relative to the VAP results. The largest differences in
$\mbox{log}T_\alpha$ are observed in the case of very heavy isotopes
($N>180$).

Similarly, Figure \ref{fig-ta-cor} displays the correlations between
the half-lives data for the $\mbox{log}(T_\alpha/y)$ versus the data
of the VAP model. The largest model discrepancies which reaches 4
orders of magnitude concern the long-living
($\mbox{log}(T_\alpha/y)>-8$) isotopes- this corresponds to the region
of superheavy nuclei which is of the main concern of both the
experiment and the theory.

\section{Summary}
We have calculated $\alpha$-decay half-lives using four pairing
models: seniority (BCSg), $\delta$-pairing (BCS$\delta$),
Lipkin-Nogami model and variation after projection method. As
expected, both the kind of pairing interaction involved and the method
of solution influence the $\alpha$-decay half-lives.
 
The largest discrepancies measured for 70 nuclei relative to the
results of the VAP method are observed in the case of Lipkin-Nogami
approach (an average deviation $\sigma_{\rm LN}=0.717\,\rm MeV$) and
BCS with constant pairing interaction, where $\sigma_{\rm
  BCSg}=0.654\,\rm MeV$.  Surprisingly, the $\delta$-pairing BCS
(BCS$\delta$) model results are closer to the variation after
projection(VAP) data in comparison to those obtained from the
Lipkin-Nogami approach. In the latter case the mean deviation from VAP
data $\sigma_\delta=0.438\,\rm MeV$.

The half-lives of long lived superheavy ($N>180$) nuclei are
determined with the largest uncertainty.

The problem of the description of the alpha decay of superheavy nuclei
is still present in the realm of superheavy game and is a challenge
for the theory to invent new, more powerful alpha decay models
describing the phenomenon.

\section{Acknowledgemens}
This work is partly supported by the Committee of Scientific Research
(KBN) of the Polish Ministry of Science and Education under contract
No.  1P03B13028 (2005-2006); by the French Embassy in Warsaw sponsoring
the stay of one of the authors (K.S.) at the University of Bordeaux
and by the Polish Ministry of Science and Education under 
Contract N~202~179~31/3920.


\end{document}